\documentstyle[preprint,floats,epsf,tighten,aps]{revtex} 

\begin{document}

\def\OMIT#1{{}}

\preprint{\vbox{ \hbox{LBNL--47551} \hbox{UTPT--01--05} \hbox{CALT--68--2316}
  \hbox{hep-ph/0103020} \hbox{} }}

\title{Comment on studying the corrections to factorization\\ 
  in $B\to D^{(*)}X$}

\author{Zoltan Ligeti,$^{a}$ Michael Luke$^{\,a,b}$ and Mark B.\ Wise$^{\,c}$}

\address{ \vbox{\vskip 0.truecm}
  $^a$Ernest Orlando Lawrence Berkeley National Laboratory \\
    University of California, Berkeley, CA 94720 \\[8pt]
  $^b$Department of Physics, University of Toronto, \\
    60 St.\ George Street, Toronto, Ontario, Canada M5S 1A7 \\[8pt]
  $^c$California Institute of Technology, Pasadena, CA 91125 }

\maketitle

\begin{abstract}%
We propose studying the mechanism of factorization in exclusive decays of the
form $B\to D^{(*)}X$ by examining the differential decay rate as a function of
the invariant mass of the light hadronic state $X$.  If factorization works
primarily due to the large $N_c$ limit then its accuracy is not expected to
decrease as the $X$ invariant mass increases.  However, if factorization is
mostly a consequence of perturbative QCD then the corrections should grow with
the $X$ invariant mass.  Combining data for hadronic tau decays and
semileptonic $B$ decays allows tests of factorization to be made
for a variety of final states.  We discuss the examples of $B\to
D^*\pi^+\pi^-\pi^-\pi^0$ and $B\to D^*\omega\pi^-$.  The mode $B\to
D^*\omega\pi^-$ will allow a precision study of the dependence of the
corrections to factorization on the invariant mass of the light hadronic state.

\end{abstract}

\newpage

An understanding of nonleptonic $B$ decays is important for the study of $CP$
violation at $B$ factories.  However, there is at present no systematic,
model-independent treatment of all such decays.  In certain kinematic
situations, factorization, according to which the matrix element of four-quark
operators can be written as the product of pairs of matrix elements of
two-quark operators, can be justified using perturbative QCD
\cite{qcdfact,bbns}.  An example is $B\to D^{(*)}X$, where $X$ denotes a
hadronic state with low invariant mass, and the $D^{(*)}$ inherits the light
degrees of freedom from the $B$ meson.  The large $N_c$ limit of QCD
\cite{tHooft} provides a different explanation of factorization, which is
independent of the kinematics of the final state.  Factorization has been shown
to be consistent with experiment in certain two-body decays, such as $B\to
D^{(*)}\pi$ and $D^{(*)}\rho$.  Currently, this represents a test of
factorization at the $10$--$20\%$ level~\cite{fact}.  Factorization has also
been observed to hold within the (presently sizable) errors in decays of the
type $B\to D^{(*)}D_s^{(*)}$, where no perturbative QCD justification has been
presented~\cite{rosner}.

In this paper we examine how the decays $B\to D^{(*)}X$, where $X$ is a final
state containing two or more light hadrons, allow us to study the corrections
to factorization\footnote{These decays have also been discussed as tests of
factorization by Reader and Isgur\cite{ReIs}.}, and help determine the role
that perturbative QCD plays.  In such decays, factorization can be studied as a
function of $m_X$, the invariant mass of the final hadronic state produced by
the light quark current.  The perturbative QCD arguments for factorization
depend on the light quarks being produced in an almost collinear state (for
this reason, it has also been argued that factorization should hold for $B\to
D^{(*)}+{\rm jet}$\cite{aglietticorbo}).  Corrections are therefore expected to
grow with $m_X/E_X$, where $E_X$ is the energy of the hadronic final state in
the $B$ rest frame, since this quantity characterizes the 
%{\it Lichtkegelartigkeit}\footnote{i.e.\ closeness to the light cone} 
%of the light hadronic state.
deviation of the hadronic state $X$ from the light cone.  
However, if factorization works primarily due to the large $N_c$ limit then its
accuracy is not expected to decrease as $m_X$ increases.  The behavior of
factorization violations with $m_X$ can therefore distinguish the perturbative
QCD explanation of factorization from that of the large $N_c$ limit.  Na\"\i
vely, the $1/N_c^2$ terms that  violate factorization are $O(10\%)$ in the
amplitude and $O(20\%)$ in the rate.  When factorization is tested with a
greater precision than this, it may be possible to see corrections to
factorization which grow with $m_X$, which would provide evidence that
perturbative QCD plays an important role in factorization.

Nonleptonic $B$ meson decays arise dominantly due to the weak Hamiltonian
\begin{equation}\label{HW}
H^{{\rm nl}} = {4G_F\over\sqrt2}\, V_{cb} V_{uq}^*\, 
  \Big[ c_1(m_b)\, (\bar c\,\gamma_\mu P_L\, b)\, (\bar q\,\gamma^\mu P_L\, u)  
  + c_2(m_b)\, (\bar q\,\gamma_\mu P_L\, b)\, (\bar c\,\gamma^\mu P_L\, u) 
  \Big]\,,
\end{equation}
where $P_L=(1-\gamma_5)/2$ and $q=d,s$.  The Hamiltonian in Eq.~(\ref{HW}) is
renormalized at the scale $\mu=m_b$.  The coefficients $c_i$ are given by 
$c_{1,2}=(c_+ \pm c_-)/2$, where
\begin{equation}\label{cpcm}
c_+(\mu) = \left[{\alpha_s(\mu) \over \alpha_s(M_W)} \right]^{-6/23},
  \qquad
c_-(\mu) = \left[{\alpha_s(\mu) \over \alpha_s(M_W)} \right]^{12/23},
\end{equation}
in the leading logarithmic approximation.  At next-to-leading order,
$c_1(m_b)=1.13$ and $c_2(m_b)=-0.29$ for $\alpha_s(M_Z)=0.118$.  Under the
factorization hypothesis, the hadronic matrix element of $H^{{\rm nl}}$
can be written as
\begin{equation}
\langle X D^{(*)} | H^{{\rm nl}} | B\rangle
  = {4G_F\over\sqrt2}\, V_{cb} V_{uq}^* 
  \left(c_1(m_b)+{1\over 3}\, c_2(m_b)\right)
  \langle D^{(*)} | \bar c\gamma_\mu P_L b  | B\rangle 
  \langle X|\bar q \gamma^\mu P_L u | 0\rangle + \dots \,,
\end{equation}
where the ellipses denote factorization violating terms.  The first matrix
element can be determined from semileptonic $B\to D^{(*)} \ell\bar\nu$ decays,
while the second can be determined from the hadronic tau decay $\tau \to X 
\nu_\tau$.  

The differential rate for tau decay to a given hadronic final state $X$ may 
be written as
\begin{equation}\label{tauspec}
{{\rm d}\Gamma(\tau\to X\nu_\tau) \over {\rm d}m_X^2}  
  = {G_F^2 |V_{ud}|^2 \over 32\pi^2\, m_\tau^3}\, 
  (m_\tau^2-m_X^2)^2\, (m_\tau^2+2m_X^2)\, v_X(m_X^2) \,,
\end{equation}
where $m_X^2$ is the invariant mass-squared of the hadronic final state $X$,
and $v_X(m_X^2)$ is related to the transverse part of the vacuum polarization
amplitude. This holds in general even if $X$ is a nonresonant multibody final
state.  The only assumption is that the weak hadronic current is conserved. 
For the vector part of the current, corrections to this assumption are
suppressed by $m_u-m_d$, while for the axial part they are suppressed by
$m_u+m_d$.  Perturbative electroweak corrections have been neglected in
Eq.~(\ref{tauspec}).

The CLEO collaboration has measured $v_X(m_X^2)$ for two-, three-, and
four-pion final states\cite{CLEOtau2pi,CLEOtau4pi}.  The two-pion final states
are quasi-two-body, being dominated by the $\rho$ resonance.  Since the $\rho$
is narrow, ${\rm d}\Gamma(B\to D^{(*)}\pi\pi) / {\rm d}m_X^2$ is a steeply
falling function of $m_X^2$ in the region we are interested in, and so is less
useful for our purposes than the three- and four-pion final states.  The
three-pion final states receive a large $a_1$ contribution, however, the $a_1$
is rather broad and there is evidence for nonresonant contributions in $\tau$
decay to three pions.  The four-pion final states are not dominated by any one
resonance.

In terms of $v_X(m_X^2)$, the factorization prediction for the decay 
$B\to D^{(*)}X$ is
\begin{equation}\label{prediction}
{ {\rm d}\Gamma(B\to D^{(*)} X) / {\rm d}m_X^2 \over 
  {\rm d}\Gamma(B\to D^{(*)} \ell\bar\nu) / {\rm d}m_X^2}
= 3\pi \left(c_1(m_b)+{c_2(m_b)\over 3} \right)^2
  v_X(m_X^2)\, (1 + \delta_{NF}) \,,
\end{equation}
where $\delta_{\rm{NF}}$ denotes the nonfactorizable contributions, and in the
denominator of the left-hand side $m_X$ denotes the dilepton invariant mass. 
In the large $N_c$ limit $\delta_{NF}$ is of order $1/N_c$, whereas in the
perturbative QCD approach it is expected to contain both perturbative
corrections and nonperturbative ones suppressed by powers of $m_X / E_X$, where
$E_X = (m_B^2 + m_X^2 - m_{D^{(*)}}^2) / (2m_B)$ is the energy of the state $X$
in the $B$ rest frame.  The ratio $m_X / E_X$ changes from $0.43$ at $m_X =
1\,$GeV to $0.70$ at $m_X = m_\tau$, so if there were order $m_X / E_X$
corrections to factorization then the accuracy of the predictions should change
significantly over the accessible range of $m_X$.

\begin{figure}
\centerline{\epsfysize=8truecm \epsfbox{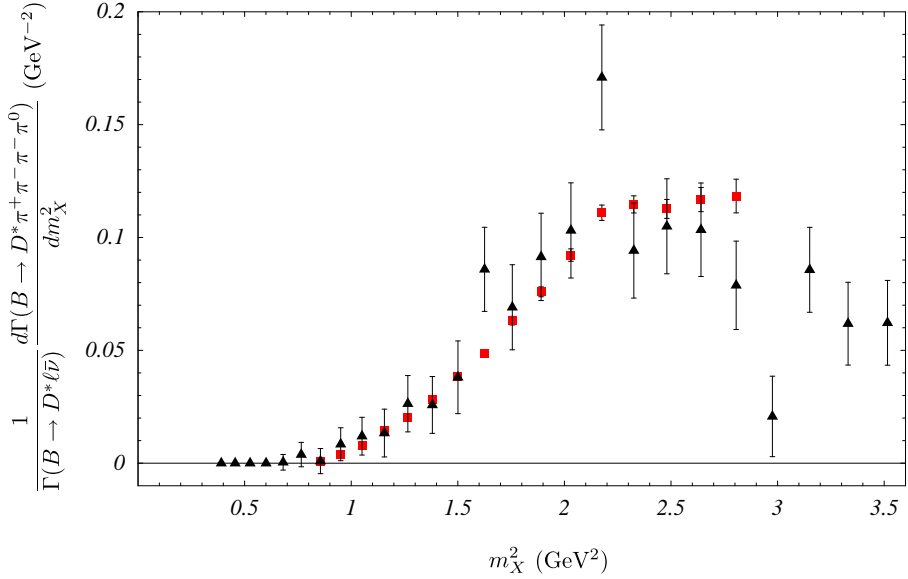}}
\caption[1]{${\rm d}\Gamma(B\to D^*\pi^+\pi^-\pi^-\pi^0)/{\rm d}m_X^2$, where 
$m_X$ is the $\pi^+\pi^-\pi^-\pi^0$ invariant mass, normalized to the 
semileptonic width $\Gamma(B\to D^*\ell\bar\nu)$.  The triangles are the $B$ 
decay data~\cite{CLEObd4pi,SSJW}, and the squares are our predictions using the 
$\tau$ data~\cite{CLEOtau4pi}.  There is an additional 9\%  
uncertainty in the $B$ decay data from the overall normalization\cite{SSJW}.}
\label{fourpifig}
\end{figure}

In Fig.~\ref{fourpifig} we plot the $m_X^2$ distribution in $B\to D^*
\pi^+\pi^-\pi^-\pi^0$ decay, normalized to the $B\to D^*\ell\bar\nu$ rate, as
predicted by Eq.~(\ref{prediction}).  We used the CLEO fit to the shape of 
${\rm d}\Gamma(B\to D^*\ell\bar\nu) / {\rm d}m_X^2$~\cite{CLEOIWfn} and the
$\tau$ decay data~\cite{CLEOtau4pi}.  This decay mode is interesting for
testing factorization because it is not dominated by a single narrow resonance,
and its branching fraction is large, almost $2\%$~\cite{CLEObd4pi}.   However,
this mode has a significant disadvantage for testing factorization because of a
potentially sizeable background from processes where one or more of the pions
are emitted from the $(\bar c b)$ current.  In this case, the $(\bar c b)$
current creates a charm state $C$ of the type $D^* + n\,\pi$ ($1 \leq n \leq
3$), and the $(\bar d u)$ current creates the remaining $4-n$ pions in the
final state.  In the formal limit when $m_b-m_c$ is large this background is
parametrically suppressed, since the light hadrons from the $C$ decay combined
with those from the $(\bar d u)$ current will have large invariant mass over
most of the phase space.  This can, however, be a sizable background for the
physical $m_b$ and $m_c$ masses.

Because of the problem of backgrounds of this type, a more promising way to
test the $m_X^2$ dependence of factorization is to focus on the subset of these
decays which proceeds via $B\to D^*\omega\pi^-$.  In this case, the backgrounds
are easier to constrain, because the $(\bar d u)$ current must create the
$\pi^-$ in the final state.  Therefore, the $(\bar c b)$ current would have to
create charmed states, $C_{(D^*\omega)}$, which can decay into $D^*\omega$. 
This background can be related using factorization (in the most reliable case
of the current only making one pion) to the $B\to C_{(D^*\omega)} \ell\bar\nu$
decay rate.  Assuming that $C_{(D^*\omega)}$ is a narrow resonance,
\begin{equation}
\Gamma(B\to C \pi) = {3\pi^2|V_{ud}|^2 f_\pi^2\over m_B\, m_C}
  \left( c_1(m_b) + \frac{c_2(m_b)}3 \right)^2 \left(
  {{\rm d}\Gamma(B\to C \ell\bar\nu)\over {\rm d}w} \right)_{w_{\rm max}} \,,
\end{equation}
where $w = v_B \cdot v_C$.  Assuming that the $C_{(D^*\omega)}$ mass is its
minimal value $m_C^{\rm min} = m_{D^*} + m_\omega$ and that semileptonic $B\to
C_{(D^*\omega)} \ell\bar\nu$ rate depends on $w$ as $\sqrt{w^2-1}$, we find
that $\Gamma(B\to C_{(D^*\omega)}\pi) \simeq 0.1 \times \Gamma(B\to
C_{(D^*\omega)}\ell\bar\nu)$.  Therefore, this background can be constrained by
measuring the $B\to D^*\omega \ell\bar\nu$ decay rate.  The measured $B\to
D^*\omega\pi^-$ branching fraction is about $0.3\%$~\cite{CLEObd4pi}.  Thus,
for example, a bound of ${\cal B}(B\to C_{(D^*\omega)}\ell\bar\nu) < 0.3\%$
would imply that the feed-down from higher mass $C_{(D^*\omega)}$ states to the
$D^*\omega\pi^-$ final state is less than $10\%$.  This is a very conservative
estimate, since the $C_{(D^*\omega)}$ branching fraction to $D^*\omega$ is
probably significantly less than unity.  Moreover, the $B\to
C_{(D^*\omega)}\ell\bar\nu$ branching fraction is probably much smaller than
$0.3\%$; this is certainly the case in most models\cite{isgw2}. Therefore, we
expect that the $B\to D^*\omega\pi^-$ decay will only receive a very small
background from higher mass charm states.

\begin{figure}
\centerline{\epsfysize=8truecm \epsfbox{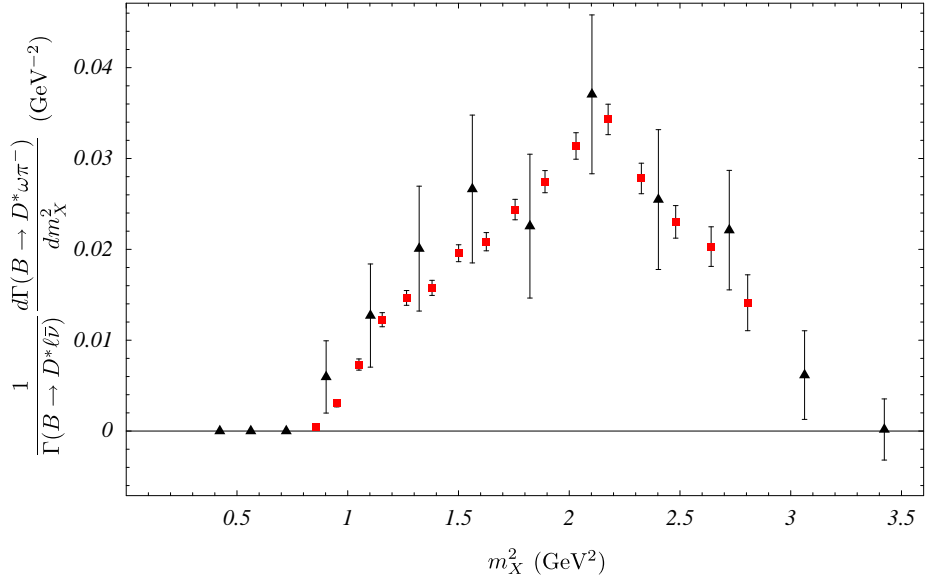}}
\caption[1]{${\rm d}\Gamma(B\to D^*\omega\pi)/{\rm d}m_X^2$, where $m_X$ 
is the $\omega\pi^-$ invariant mass, normalized to the semileptonic width 
$\Gamma(B\to D^*\ell\bar\nu)$.  The triangles are the $B$ decay 
data~\cite{CLEObd4pi}, and the squares are our predictions using the $\tau$ 
data~\cite{CLEOtau4pi}. There is an additional 9\%  
uncertainty in the $B$ decay data from the overall normalization\cite{SSJW}.}
\label{omegapifig}
\end{figure}

In Fig.~\ref{omegapifig} we plot the $m_X^2$ distribution in $B\to
D^*\omega\pi^-$ decay, normalized to the $B\to D^*\ell\bar\nu$ rate, as
predicted by Eq.~(\ref{prediction}) and the $\tau$ decay
data~\cite{CLEOtau4pi}.  This decay receives a large contribution from a
$\rho'$ resonance with mass around $1.4-1.5\,$GeV~\cite{CLEOtau4pi,CLEObd4pi}. 
However, the $\rho'$ is very broad, with a width estimated to be around
$400-500\,$MeV.  Since we would like to make a precision test of factorization
and learn about the size of corrections suppressed by powers of $m_X/E_X$ and
$\alpha_s$, one needs tests that do not make assumptions based on resonance
models or heavy quark symmetry~\cite{book}, which may have corrections of
similar size.  This is best achieved by looking at the differential mass
distribution, which should also teach us about how factorization breaks down
(if it does at all) at higher masses.  

Controlling this background in the $B\to D^* \pi^+\pi^-\pi^-\pi^0$ decay is
more complicated than for $B\to D^*\omega\pi^-$, and may make it difficult to
make a precise study of factorization in this channel.  The contribution when
$C$ contains two or three pions is expected to be small and can be constrained
by bounding the semileptonic rates as in the case of $B\to D^*\omega\pi^-$. 
The contribution when one pion comes from the $(\bar c b)$ current and three
pions from the $(\bar d u)$ current may be estimated from the measured
$v_{3\pi}(m_X^2)$ and the $B\to D^*\pi\ell\bar\nu$ branching fraction.  Using
the ALEPH data~\cite{aleph} for $v_{3\pi}(m_X^2)$ measured in $\tau$ decay and
${\cal B}(B\to D^*\pi\ell\bar\nu) \sim 1\%$, it appears that this background
may be significant.  The size of this background can be tested experimentally
by measuring the $B\to D^{*0} \pi^+\pi^+\pi^-\pi^-$ decay rate, since this
decay only receives contributions from higher mass charm states which decay
into a $D^{*0}$ plus one or three pions.  Unless the $B\to D^{*0}
\pi^+\pi^+\pi^-\pi^-$ rate is much smaller than the $B\to D^*
\pi^+\pi^-\pi^-\pi^0$ rate, it will be difficult to make a precise test of
factorization in $B\to D^* \pi^+\pi^-\pi^-\pi^0$.

Data on $e^+ e^-\to\ $hadrons can also be used to determine $v_X(m_X^2)$ and
predict the mass spectrum in $B\to D^{(*)}X$ decay~\cite{GiMi}.  Currently the
CMD--2 data on $e^+e^- \to 4\pi$ are available up to $m_X^2 = (1.4\,
\rm{GeV})^2$~\cite{CMD2}; measurements of this cross section at higher energies
would allow testing factorization at values of $m_X^2 > m_\tau^2$.

In summary, for a precision study of corrections to factorization in $B\to
D^{(*)}X$ decays (where $X$ contains only light hadrons), it is best not to
make assumptions based on heavy quark symmetry or resonant decomposition
(especially for broad states).  Rather, one should use the differential $m_X^2$
distribution with a fixed hadron content and compare with the predicted rate
based on factorization using the spectral function extracted from $\tau$ decay.
Over the accessible kinematic range the parameter $m_X / E_X$, which controls
the accuracy of the predictions if the primary reason for factorization is
perturbative QCD, changes from about $0.4$ to $0.7$, making $m_X$-dependent
violations of factorization potentially observable. In this paper we
concentrated on final states containing four pions, but this approach is
general and can be applied to test factorization in decays to any light
hadronic final state where backgrounds from the $(\bar c b)$ current producing
some of the light hadrons can be controlled.  Studying the available data in
$B\to D^* \pi^+\pi^-\pi^-\pi^0$ and $B\to D^*\omega\pi^-$ decays, we found no
evidence of factorization becoming a worse approximation at higher values of
$m_X$.  It will be very interesting to make a precision study, which should
be possible at the $B$ factories.  The decay mode $B\to D^*\omega\pi^-$ is
particularly well suited for this purpose.

\acknowledgements

We thank Sheldon Stone, Jianchun Wang and Alan Weinstein for help
with the CLEO data and discussions.
Z.L.\ was supported in part by the Director, Office of Science, 
Office of High Energy and Nuclear Physics, Division of High Energy Physics, 
of the U.S.\ Department of Energy under Contract DE-AC03-76SF00098.
M.L.\ was supported in part by the Natural Sciences and Engineering
Research Council of Canada.  M.B.W.\ was supported in part by the 
Department of Energy under Grant No.~DE-FG03-92-ER40701.


\begin{references}


\bibitem{qcdfact}
J.D. Bjorken, Nucl. Phys. Proc. Suppl. 11 (1989) 325; \\
M.J. Dugan and B. Grinstein, Phys. Lett. B255 (1991) 583; \\
H.D. Politzer and M.B. Wise, Phys. Lett. B257 (1991) 399.

\bibitem{bbns}
M. Beneke, G. Buchalla, M. Neubert and C.T. Sachrajda,
Phys. Rev. Lett. 83 (1999) 1914;
Nucl. Phys. B591 (2000) 313.

\bibitem{tHooft}
G. 't Hooft, Nucl. Phys. B72 (1974) 461; Nucl. Phys. B75 (1974) 461.

\bibitem{fact}
J.L. Rosner, Phys. Rev. D42 (1990) 3732; \\
D. Bortoletto and S. Stone, Phys. Rev. Lett. 65 (1990) 2951; \\
M. Neubert and B. Stech, in {\it Heavy Flavours II} 
(Buras, A.J. and Lindner, M., eds.), 294-344 (hep-ph/9705292).

\bibitem{rosner}
Z. Luo and J.L. Rosner, hep-ph/0101089.

\bibitem{aglietticorbo}
U.~Aglietti and G.~Corb\`o, Phys. Lett. B431 (1998) 166.

\bibitem{ReIs}
C. Reader and N. Isgur, Phys. Rev. D47 (1993) 1007.

\bibitem{CLEOtau2pi}
F. Anderson {\it et al.}, CLEO Collaboration, Phys. Rev. D61 (2000) 112002.

\bibitem{CLEOtau4pi}
K.W. Edwards {\it et al.}, CLEO Collaboration, Phys. Rev. D61 (2000) 072003.

\bibitem{CLEOIWfn}
J.P. Alexander {\it et al.}, CLEO Collaboration, CLEO CONF 00--3.

\bibitem{CLEObd4pi}
M. Artuso {\it et al.}, CLEO Collaboration, CLEO CONF 00--1.

\bibitem{SSJW}
S. Stone and J. Wang, private communication.

\bibitem{isgw2} 
N. Isgur, D. Scora, B. Grinstein and M.B. Wise, Phys. Rev. D39 (1989) 799; \\
D. Scora and N. Isgur, Phys. Rev. D52 (1995) 2783.

\bibitem{book}
A.V. Manohar and M.B. Wise, {\it Heavy quark physics},
Cambridge Monographs on Particle Physics, Nuclear Physics, and Cosmology, 
Vol. 10.

\bibitem{aleph}
R. Barate {\it et al.}, ALEPH Collaboration, Eur. Phys. Journal C4 (1998) 409.

\bibitem{GiMi}
F.J. Gilman and D.H. Miller, Phys. Rev. D17 (1978) 1846.

\bibitem{CMD2}
N.I. Root [for the CMD-2 Collaboration], Nucl. Phys. A675 (2000) 341.


\end{references}
\end{document}